# Sb-NQR probe for superconducting property in the Pr-based filled skutterudite compound $PrRu_4Sb_{12}$

M. Yogi[1], H. Kotegawa[1], Y. Imamura[1], G. -q. Zheng[1], Y. Kitaoka[1], H. Sugawara[2], H. Sato[2]
[1]*Department of Physical Science, Graduate School of Engineering Science, Osaka University, Osaka 560-8531, Japan and*
[2]*Graduate School of Science, Tokyo Metropolitan University,*
*Minami-Ohsawa 1-1, Hachioji, Tokyo 192-0397, Japan*

We report the electronic and superconducting properties in the Pr-based filled-skutterudite superconductor $PrRu_4Sb_{12}$ with $T_c = 1.3$ K via the measurements of nuclear-quadrupole-resonance (NQR) frequency $\nu_Q$ and nuclear-spin-lattice-relaxation time $T_1$ of Sb nuclei. The temperature dependence of $\nu_Q$ has revealed the energy scheme of $Pr^{3+}$ crystal electric field (CEF) that is consistent with an energy separation $\Delta_{CEF} \sim 70$K between the ground state and the first-excited state. In the normal state, the Korringa relation of $(1/T_1T)_{Pr}$=const. is valid, with $[(1/T_1T)_{Pr}/(1/T_1T)_{La}]^{1/2} \sim 1.44$ where $(1/T_1T)_{La}$ is for $LaRu_4Sb_{12}$. These results are understood in terms of a conventional Fermi liquid picture in which the Pr-$4f^2$ state derives neither magnetic nor quadrupolar degrees of freedom at low temperatures. In the superconducting state, $1/T_1$ shows a distinct coherence peak just below $T_c$, followed by an exponential decrease with a value of $2\Delta/k_BT_c = 3.1$. These results demonstrate that $PrRu_4Sb_{12}$ is a typical weak-coupling *s*-wave superconductor, in strong contrast with the heavy-fermion superconductor $PrOs_4Sb_{12}$ that is in an unconventional strong coupling regime. The present study on $PrRu_4Sb_{12}$ highlights that the Pr-$4f^2$derived non-magnetic doublet plays a key role in the unconventional electronic and superconducting properties in $PrOs_4Sb_{12}$.

PACS numbers: 71.27.+a, 76.60.-k

Filled-skutterudite compounds $ReT_4Pn_{12}$ (Re = rare earth; T = Fe, Ru and Os; Pn = pnictogen) show rich properties. $PrRu_4P_{12}$ and $PrFe_4P_{12}$ show a metal-insulator transition and undergo into an anomalous heavy-fermion (HF) state, respectively, whereas $PrRu_4As_{12}$, $PrRu_4Sb_{12}$ and $PrOs_4Sb_{12}$ exhibit a superconducting (SC) transition.[1,2,3,4] Bauer *et al.* reported that $PrOs_4Sb_{12}$ shows HF behavior and superconducts at $T_c = 1.85$ K. It is the first Pr-based HF superconductor.[4] Its HF state was inferred from the jump in the specific heat at $T_c$, the slope of the upper critical field $H_{c2}$ near $T_c$, and the electronic specific-heat coefficient $\gamma \sim 350 - 500$ mJ/mole K$^2$. Magnetic susceptibility, thermodynamic measurements, and inelastic neutron scattering experiments revealed the ground state of the $Pr^{3+}$ ions in the cubic crystal electric field (CEF) to be the $\Gamma_3$ nonmagnetic doublet[4,5]. In the Pr-based compounds with the $\Gamma_3$ ground state, electric quadrupolar interactions play an important role. In analogy with a quadrupolar Kondo model,[6] it was suggested that the HF-like behavior exhibited by $PrOs_4Sb_{12}$ may have something to do with a Pr-$4f^2$-derived quadrupolar Kondo-lattice. An interesting issue to be addressed is what role of Pr-$4f^2$-derived quadrupolar fluctuations plays in relevance with the onset of the superconductivity in this compound.

Meanwhile, Kotegawa *et al.* have reported the Sb-NQR results which evidence the HF behavior and the unconventional SC property in $PrOs_4Sb_{12}$.[7] The temperature ($T$) dependencies of nuclear-spin-lattice-relaxation rate, $1/T_1$ and nuclear-quadrupole-resonance (NQR) frequency unraveled a low-lying CEF splitting below $T_0 \sim 10$ K, associated with the $Pr^{3+}(4f^2)$-derived ground state. The analysis of $T_1$ suggests the formation of HF state below $\sim 4$ K. In the SC state, $1/T_1$ shows neither a coherence peak just below $T_c = 1.85$ K nor a $T^3$ like power-law behavior observed for *anisotropic* HF superconductors with line-node gap. An *isotropic* energy-gap with $\Delta/k_B = 4.8$ K is suggested to open up already below $T^* \sim 2.3$ K. It is surprising that $PrOs_4Sb_{12}$ looks like an *isotropic* HF superconductor – it may indeed argue for Cooper pairing via quadrupolar fluctuations. Also, $PrRu_4Sb_{12}$ was reported to undergo a SC transition at $T_c$= 1.3 K from the measurements of the electrical resistivity and specific heat as well as $LaRu_4Sb_{12}$ with $T_c$=3.58 K.[3] It can be informative to compare $PrRu_4Sb_{12}$ with $PrOs_4Sb_{12}$ and the related La-based superconductors as shown in Table I.[8]

The localized character of $4f$ electrons, namely the closeness of the respective Fermi surfaces with those in $LaRu_4Sb_{12}$ and $LaOs_4Sb_{12}$, has been confirmed in $PrRu_4Sb_{12}$ and $PrOs_4Sb_{12}$ based on the de Haas-van Alphen (dHvA) experiment.[8,13] On the contrary, the mass enhancement in $PrRu_4Sb_{12}$ is much smaller than in $PrOs_4Sb_{12}$. For $PrOs_4Sb_{12}$, the CEF ground state was inferred to be the non-Kramers $\Gamma_3$ doublet carrying quadrupole moments, whereas the ground state for $PrRu_4Sb_{12}$ to be the $\Gamma_1$ singlet.[3,9] Recently, however, there are several reports that are consistent with the CEF ground state for $PrOs_4Sb_{12}$ being the $\Gamma_1$ singlet.[10,11,12] On the comparison in $T_c$ with the La compounds, the two compounds have different trend; $T_c$ for $PrOs_4Sb_{12}$ is higher than that for La compounds, which is unusual if we take into account that $PrOs_4Sb_{12}$ contains the magnetic element Pr ion. These remarkable differences in the underlying CEF level scheme and hence electronic and SC characteristics between $PrOs_4Sb_{12}$ and $PrRu_4Sb_{12}$ may be ascribed to an intimate change in the hybridization strength of Pr-4f state with conduction electrons comprising of respective $Os_4Sb_{12}$- and $Ru_4Sb_{12}$-cage. In this context, it is needed that further light is shed upon the SC and electronic characteristics in the Pr-based superconductors.

In this paper, we report the normal and SC properties in the filled-skutterudite compound $PrRu_4Sb_{12}$ and $LaRu_4Sb_{12}$ via the measurements of NQR frequency $\nu_Q$ and nuclear-spin-lattice-relaxation time $T_1$ of Sb nuclei.

Single crystals of $PrRu_4Sb_{12}$ and $LaRu_4Sb_{12}$ were grown by the Sb-flux method.[3] The observed dHvA oscillations in both compounds confirm the high quality of the samples.[13] Measurement of ac-susceptibility confirmed the SC transitions at $T_c$ = 1.3 K and 3.5 K for $PrRu_4Sb_{12}$ and $LaRu_4Sb_{12}$, respectively. The single crystal was crushed into powder for Sb-NQR measurement. The $^{121,123}$Sb-NQR measurements were performed using the conventional saturation-recovery method at zero field ($H = 0$). The NQR-$T_1$ measurement was carried out using the NQR transition $2\nu_Q$ at the $T$ range of $T$=0.24 K - 240 K using a $He^3$-$He^4$ dilution refrigerator.

Fig.1(a) displays the $^{121,123}$Sb-NQR spectra at 4.2 K. Sb nuclei has two isotopes $^{121}$Sb and $^{123}$Sb. The respective nuclear spin $I$=5/2 ($^{121}$Sb) and 7/2 ($^{123}$Sb) have natural abundance 57.3 and 42.7%, and nuclear gyromagnetic ratio $\gamma_N$=10.189 and 5.5175 [MHz/T], giving rise to two and three NQR transitions, respectively. Fig.1(b) indicates the $T$ dependencies of $\nu_Q(T)$ derived from the $^{123}$Sb-$2\nu_Q$ transition in $PrRu_4Sb_{12}$ and $LaRu_4Sb_{12}$. The inset indicates $\delta\nu_Q(T) = \nu_Q(T)_{Pr} - \nu_Q(T)_{La}$, which subtracts the common effect due to lattice expansion in the both compounds. $\nu_Q(T)$ reveals a progressive increase upon cooling below $T \sim 70$ K, which is considered to be due to the CEF splitting. Note, as shown in Fig.1(c), that the $\delta\nu_Q(T) = \nu_Q(T)_{Pr} - \nu_Q(T)_{La}$ in $PrOs_4Sb_{12}$ was observed to be increased below a temperature comparable to the CEF splitting $\Delta_{CEF} \sim$10 K between the ground state and the first excited state. From this comparison, $\Delta_{CEF} \sim 70$ K is expected in $PrRu_4Sb_{12}$. This is almost consistent with the analysis of susceptibility and resistivity.[3,9]

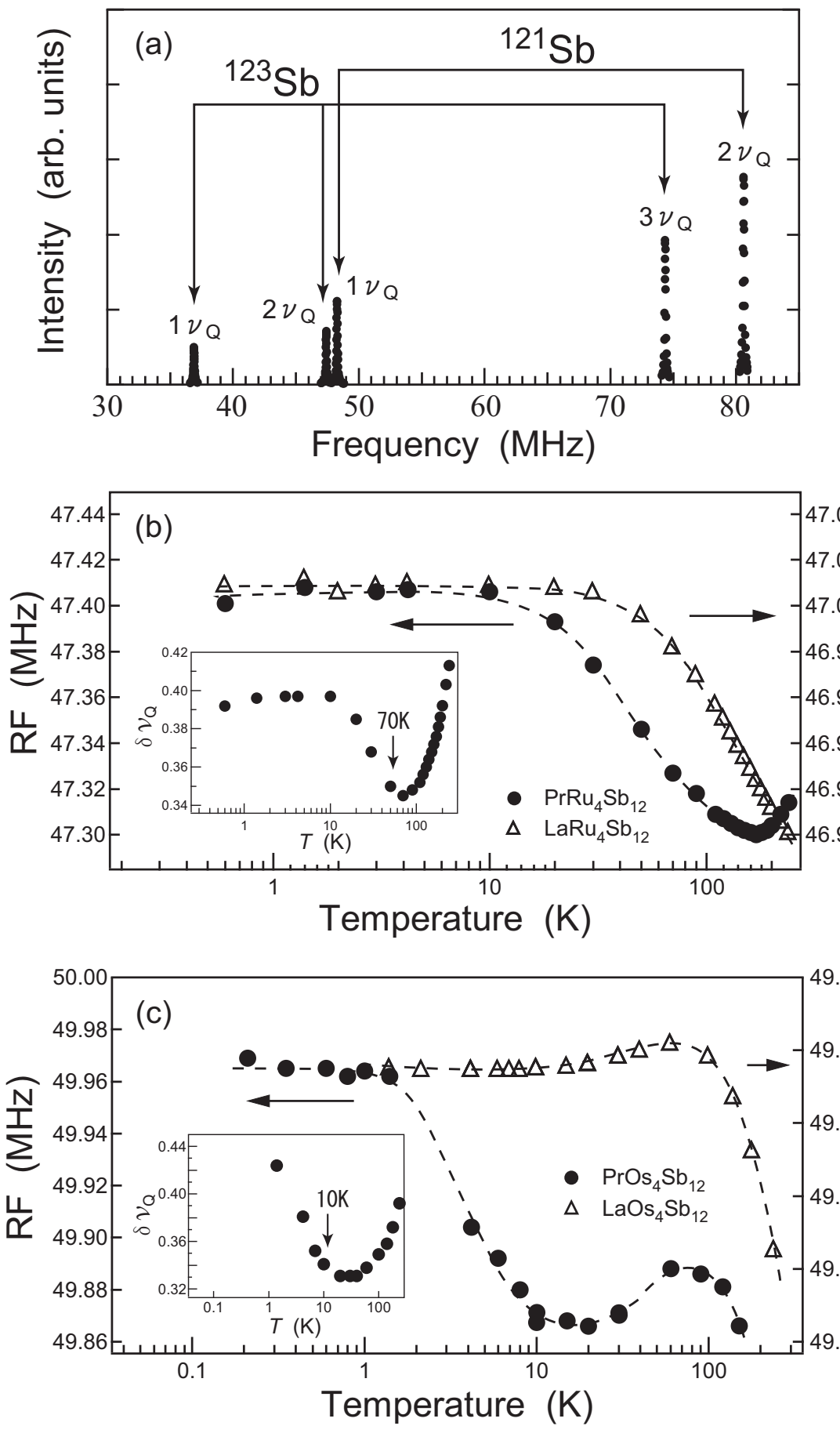

FIG. 1: (a) $^{121}$Sb- and $^{123}$Sb-NQR spectra in $PrRu_4Sb_{12}$. (b) The temperature dependence of NQR frequency $\nu_Q$ for $PrRu_4Sb_{12}$ and $LaRu_4Sb_{12}$ at the $^{123}$Sb-$2\nu_Q$ transitions. The inset indicates the Pr-derived contribution in $\nu_Q$, $\delta\nu_Q = (\nu_Q)_{Pr} - (\nu_Q)_{La}$. (c) $T$ dependence of NQR frequency for $PrOs_4Sb_{12}$ and $LaOs_4Sb_{12}$ at $^{123}$Sb-$2\nu_Q$ transitions.[7] The Inset indicates $\delta\nu_Q = (\nu_Q)_{Pr} - (\nu_Q)_{La}$..

Fig.2 presents the $T$ dependencies of $(1/T_1T)$ for $PrRu_4Sb_{12}$ and $LaRu_4Sb_{12}$. In the normal state, $T_1$ reveals a Korringa relation $(1/T_1T)_{Pr} = 1.73(s{\cdot}\mathrm{K})^{-1}$ for $PrRu_4Sb_{12}$, being comparable to $(1/T_1T)_{La} = 1.2(s{\cdot}\mathrm{K})^{-1}$ for $LaRu_4Sb_{12}$. The $1/T_1T$ =const. law deviates at temperatures higher than $\sim$ 30K in $PrRu_4Sb_{12}$. Since such a deviation is seen in $LaRu_4Sb_{12}$ above $\sim$ 25K as well, these deviations are not derived by the presence of $Pr^{3+}$ ions, but may be ascribed to a conduction-band derived effect inherent to the filled-skutterudite structure. In the filled-skutterudite structure, a Pr atom forms in a body centered cubic structure, surrounded by a cage of corner-sharing $Ru_4Sb_{12}$ octahedra. The cage might begin to stretch with increasing $T$. This stretching motion of cage may be relevant to the decrease in a value of $1/T_1T$=const. for $PrRu_4Sb_{12}$ and $LaRu_4Sb_{12}$ and $LaRu_4P_{12}$.[14] The measurements of the dHvA effect and the electronic specific heat for $PrRu_4Sb_{12}$ and $LaRu_4Sb_{12}$ revealed that the mass-renormalization effect in the Fermi liquid state is not so significant in $PrRu_4Sb_{12}$, suggesting that $Pr^{3+}$-$4f^2$ electrons are well localized in $PrRu_4Sb_{12}$. Note that the value of $1/T_1T$ is proportional to the square of the density of states $N(E_F)$ at the Fermi level. Also, it is scaled to a $T$-linear electronic contribution $\gamma$ of specific heat, giving rise to the relation of $(1/T_1T)^{1/2} \propto \gamma$. Therefore, the change in value of $(1/T_1T)^{1/2}$ is directly related to a change of $N(E_F)$ in systems . Corroborated by the fact that the value of $1/T_1T$ in $PrRu_4Sb_{12}$ is not so enhanced than that in

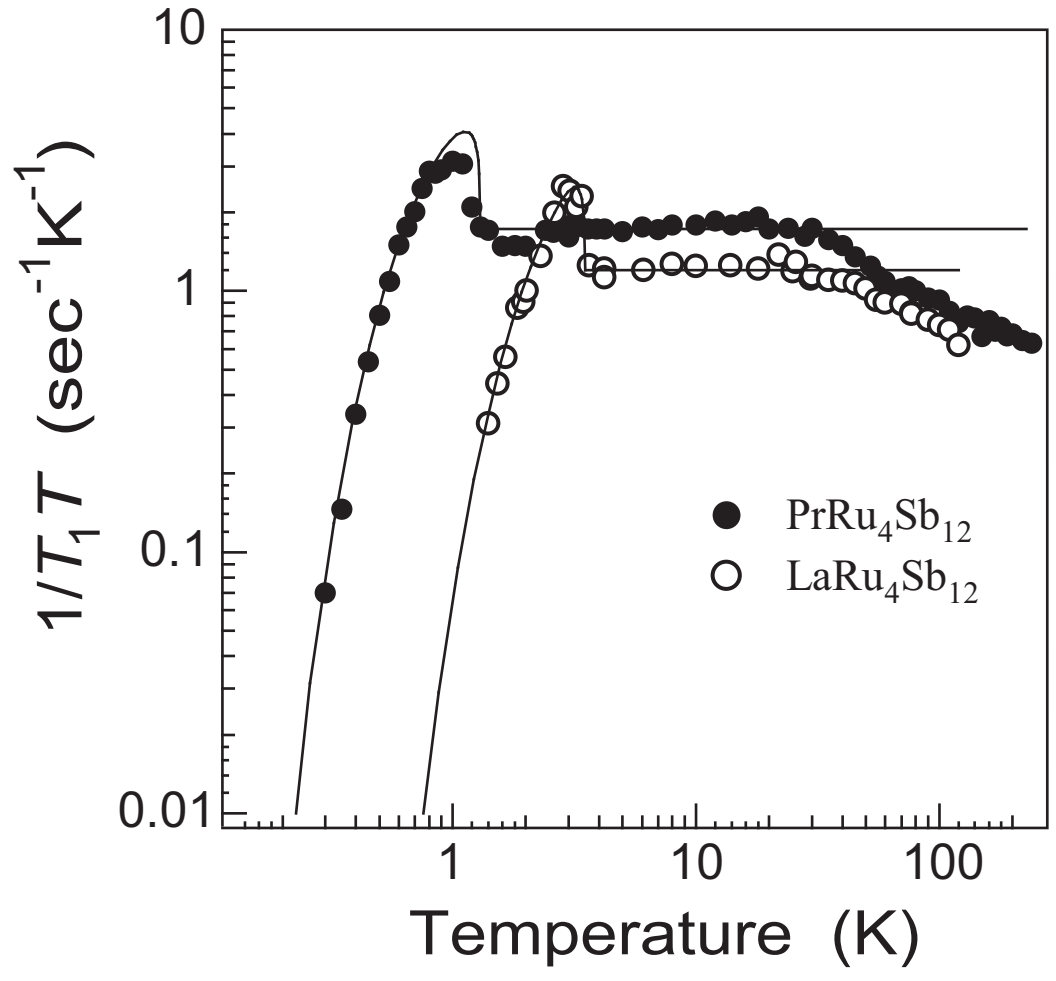

FIG. 2: Temperature dependence of $1/T_1T$ for $PrRu_4Sb_{12}$ and $LaRu_4Sb_{12}$. Solid lines are the fits calculated based on the weak-coupling $s$-wave model assuming a size of isotropic gap $2\Delta/k_BT_c$=3.1 and 3.6 for $PrRu_4Sb_{12}$ and $LaRu_4Sb_{12}$, respectively.

$LaRu_4Sb_{12}$ with a ratio of $[(1/T_1T)_{Pr}/(1/T_1T)_{La}]^{1/2} = 1.44$, we remark that the $Pr^{+3} - 4f^2$ electrons with $\Gamma_1$ singlet as the ground state does not play a vital role for electronic and magnetic properties at low temperatures in $PrRu_4Sb_{12}$.

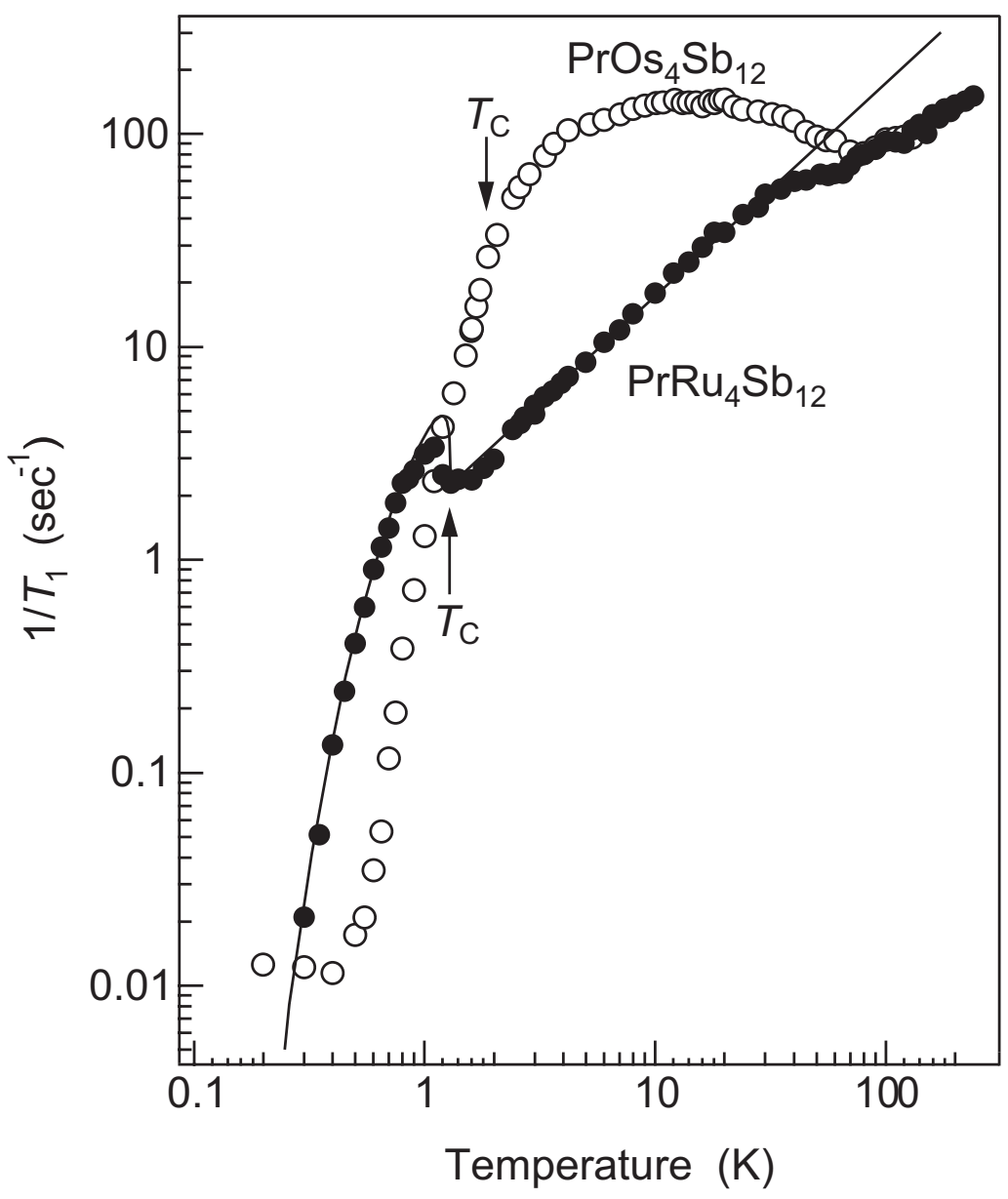

FIG. 3: Temperature dependencies of $1/T_1$ for $PrRu_4Sb_{12}$ and $PrOs_4Sb_{12}$.[7] The solid line for $PrRu_4Sb_{12}$ is the fit based on the weak-coupling $s$-wave model with $2\Delta/k_BT_c = 3.1$.

In the SC state for $PrRu_4Sb_{12}$ and $LaRu_4Sb_{12}$, $1/T_1$ shows a distinct coherence peak, followed by an exponential decrease below $T_c$ with an isotropic gap $2\Delta/k_BT_c = 3.1$ and 3.6, respectively. These results demonstrate that $PrRu_4Sb_{12}$ and $LaRu_4Sb_{12}$ are typical weak-coupling $s$-wave superconductors. In Fig.3 are shown the $T$ dependencies of $1/T_1$ for $PrRu_4Sb_{12}$ and $PrOs_4Sb_{12}$. From the comparison in the normal and SC states, it is clear that remarkable differences arise because the quadrupole degree of freedom plays vital role in $PrOs_4Sb_{12}$, associated with the $Pr^{+3}$-$4f^2$ derived non-Kramers doublet. It may indeed argue for Cooper pairing via quadrupolar fluctuations.

To summarize, the electronic and superconducting properties in the Pr-based filled-skutterudite superconductor $PrRu_4Sb_{12}$ with $T_c = 1.3$ K were investigated through the measurements of nuclear-quadrupole-resonance (NQR) frequency $\nu_Q$ and nuclear-spin-lattice-relaxation time $T_1$ of Sb nuclei. The $T$ dependence of $\nu_Q$ has revealed the energy scheme of crystal electric field (CEF) of $Pr^{3+}$ ion that is consistent with an energy separation $\Delta_{CEF} \sim 70$K between the ground state and the first-excited level. In the normal state, the Korringa relation of $(1/T_1T)_{Pr}$=const. is valid, revealing a comparable value $[(1/T_1T)_{Pr}/(1/T_1T)_{La}]^{1/2} \sim 1.44$ with $(1/T_1T)_{La}$ for $LaRu_4Sb_{12}$. These results are understood in terms of the conventional Fermi-liquid picture in which the Pr-$4f^2$ state derives neither magnetic nor quadrupolar degrees of freedom at low temperatures. In the SC state, $1/T_1$ shows a distinct coherence peak just below $T_c$, followed by an exponential decrease with the value of $2\Delta/k_BT_c = 3.1$. These results demonstrate that $PrRu_4Sb_{12}$ is a typical weak-coupling s-wave superconductor, in strong contrast with the heavy-fermion superconductor $PrOs_4Sb_{12}$ that is in a unconventional strong coupling regime.[7] The present study on $PrRu_4Sb_{12}$ highlights that the Pr-$4f^2$derived non-magnetic doublet plays a key role for the unconventional electronic and superconducting properties in $PrOs_4Sb_{12}$.

We thank Y. Aoki and H. Harima for helpful discussions. This work was supported by the COE Research (10CE2004) in Grant-in-Aid for Scientific Research from the Ministry of Education, Sport, Science and Culture of Japan. One of the author ( H. K. ) has been supported by JSPS Research Fellowships for Young Scientists.

TABLE I: Comparison of the superconducting critical temperature $T_c$, superconducting specific heat jump $\Delta C$ divided by $T_C$ ($\Delta C/T_C$). Sommerfeld coefficient, and effective mass $m_c^*$ in $RT_4Sb_{12}$ (R=La, Pr, T=Ru, Os).[8]

| | $PrOs_4Sb_{12}$ | $LaOs_4Sb_{12}$ | $PrRu_4Sb_{12}$ | $LaRu_4Sb_{12}$ |
|---|---|---|---|---|
| $T_C$ (K) | 1.85 | 0.74 | 1.3 | 3.58 |
| $\Delta C/T_C$ (mJ/K$^2$ mol) | 500 | 84 | 110 | 82 |
| Sommerfeld coefficient (mJ/K$^2$ mol) | 350∼750 | 36, 56 | 59 | 37 |
| $m_c^*/m_0$ for $\gamma$-branch | 7.6 | 2.8 | 1.6 | 1.4 |